\documentclass[prb,nofootinbib,twocolumn,superscriptaddress]{revtex4} 

\usepackage{graphicx}
\usepackage{dcolumn}
\usepackage{bm}
\usepackage{threeparttable}
\usepackage{times}
\usepackage{mathptmx}
\usepackage{lscape}
\usepackage{natbib}
\usepackage{amsmath}
\usepackage{amssymb}
\usepackage{braket}
\usepackage{color}

\def\degree{\kern-.2em\r{}\kern-.3em}

\begin{document}


\title{ Extended Configurational Polyhedra Based on Graph Representation for Crystalline Solids }

\author{Koretaka Yuge}
\affiliation{
Department of Materials Science and Engineering,  Kyoto University, Sakyo, Kyoto 606-8501, Japan\\
}%

\begin{abstract}
{ We propose theoretical approach based on combination of graph theory and generalized Ising model (GIM), which enables systematic determination of extremal structures  for crystalline solids without any information about interactions or constituent elements. The conventional approach to find such set of structure typically employs configurational polyhedra (CP) on configuration space based on GIM description, whose vertices can always be candidates to exhibit maximum or minimum physical quantities. We demonstrate that the present approach can construct extended CP whose vertices not only include those found in conventional CP, but also include other topologically and/or configurationally characteristic structures on the same dimensional configuration space with the same set of figures composed of underlying lattice points, which therefore has significant advantage over the conventional approach. 

  }
\end{abstract}


\maketitle

\section{Introduction}
In classical systems where physical quantity (especially, dynamical variables) can be a linear map for structures,  a set of special structure that has maximum or minimum physical quantity should always be restricted by spatial constraint on constituents of given system. This can be quantitatively determined by the landscape of so-called \textit{configurational polyhedra} (CP),\cite{cp} which determines maximum and minimum value that basis functions for describing structures (typically, generalized Ising model\cite{ce} (GIM) descrpition is employed) can take: Such special structures always locate on vertices of the CP. 
Since the landscape of the CP can be determined without any information about energy or constituents, we can a priori know a set of candidate structure to exhibit extremal physical quantities when condition of spatial constraint on constituents of the system is given. Using this characteristics, efficient prediction of alloy ground-state structures have amply been investigated so far, and the concept of the CP has recently been extended to finite-temperature properties,\cite{ycp} where configurational density of states (CDOS) along CP-based special coordination well characterize temperature dependence of internal energy near order-disorder phase transition. 

The practical problem for using CP is that number of its vertices exponentially increases at high-dimensional configuration space considered. It is thus fundamentally important to find out as many extremal structures as possible at low-dimensional space based on geometrically low-dimensional figures, such as symmetry-nonequivalent pairs on lattice. However, it has been shown\cite{cp} that a set of structure at vertices of CP is invariant with linear transformation of the coordination within given subspace, which directly means that when we would like to further figure out extremal structures based on the conventional CP approach, we should explicitly include information about longer-range and/or higher-dimensional figures on lattice, but such CP projected onto desired low-dimensional space typically lose significant information about vertices that are originally found at low-dimensional space based on low-dimensional figures. In order to overcome this problem in GIM-based conventional CP, the present study proposes construction of CP based on extendend graph representation unified with GIM-based description for crystalline solids, whose spectrum not only contains information about GIM pair correlations, but also includes higher-dimensional figures (or links) consisting of the corresponding pairs. We will demonstrate that the proposed CP retains characteristic vertices found for conventional CP, and also can figure out other special structures in terms of graph, on the same low-dimensional space considered.

\section{Methodology}
Very recently, we successfully construct unified representation of microscopic structure on periodic lattice (i.e., atomic configuration) in terms of both GIM and graph description. 
Explicit relationship between GIM and graph representation for given figure $R$ on given lattice is expressed by\cite{yg}
\begin{eqnarray}
\label{eq:g}
\Braket{\psi \left(\vec{\sigma}\right)}_R = \rho_R^{-1} \sum_{m} C_m\cdot \textrm{Tr} \left[ \left(\sum _{l\in R} \bm{B}_l\left(\vec{\sigma}\right)\pm {}^t\!\bm{B}_l\left(\vec{\sigma}\right) \right)^{N_R} \right], 
\end{eqnarray}
where left-hand side corresponds to multisite correlation for figure $R$ in GIM descripition. Matrix $\bm{B}$ denotes upper triangular part of extended graph Laplacian, defined by
\begin{eqnarray}
\label{eq:lp}
\bm{B}_R^{\left(\alpha\right)}\left(i,j\right) = \begin{cases} 
\sqrt{\phi_p\left(\sigma_i\right)\phi_q\left(\sigma_j\right)} & \left(p,q \in \alpha, i,j\in R, i < j\right),  \\
0 & (otherwise)
\end{cases}
\end{eqnarray}
where $p$ and $q$ denotes conventional GIM basis function index on each lattice point $i$ and $j$, and $\sigma_k$ is the pseudo-spin variable to specify atomic occupation. 
$\bm{A} = \bm{B} + {}^t\!\bm{B}$ corresponds to adjacency matrix.
 $N_R$ is the dimension of figure, $C_m$ is integer depending on the type of figure, and $\rho_R$ denotes number of the possible path to construct considered figure $R$. Here, we construct graph Laplacian (and adjacency matrix) composed of symmetry-equivalent neighboring edges: Therefore, summation in the right-hand side is taken over possible pair figure $l$ that corresponds to subfigure of figure $R$.  
From Eq.~(\ref{eq:g}), we have shown that information about higher-order links (or multisite correlations for higher-dimensional figure) in the structure can be explicitly included in the landscape of graph spectrum composed of the corresponding pair figures. 
It is thus naturally expected that when we construct CP based on the extended graph representation, we would obtain additional atomic configurations on vertices of the CP compared with GIM-based CP. 
From Eq.~(\ref{eq:lp}), it is clear that adjacency matrix for any given figure $R$ is traceless. Therefore, in order to characterize the landscape of the resultant graph spectrum for $\bm{A}$s, we here employ graph energy that is the sum of absolute of all eigenvalues for graph spectrum. Graph energy has been extensively investigated to characterize such as regularity by determining upper and/or lower bound of its value, and corresponding application has been done for molecules to relate to its energetics.\cite{ge1,ge2,ge3,ge4} 

In the present study, we prepare examples of all possible atomic configurations for A-B binary system on $4\times 4\times 4$ expansion of fcc conventional unit cell having minimal unit consisting of up 16 atoms, and calculate corresponding graph spectrum, where compositions $x$ (A$_x$B$_{1-x}$) of 0.5 and 0.75 are considered. Spin variable at each lattice point, $\sigma_i$, takes +1 (0) for occupation of A (B) element. 
For these atomic configurations, we calculate second-order moment $\mu_2$ of $\bm{A}$ for up to 3rd nearest-neighbor (3NN) pair (i.e., GIM pair correlations up to 3NN pair) to construct conventional CP, and also calculate graph energy for linear combination of the same set of $\bm{A}$ above. 


\section{Results and Discussions}
\begin{figure}[h]
\begin{center}
\includegraphics[width=1.00\linewidth]{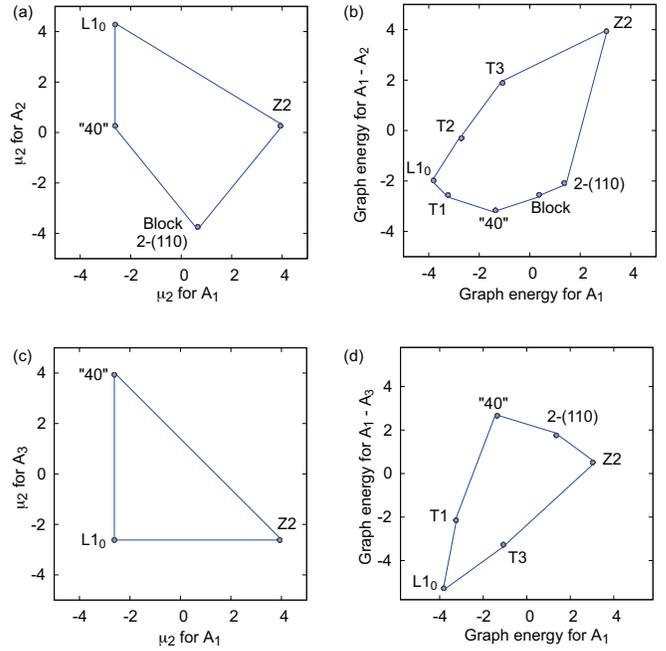}
\caption{Configurational polyhedra (CP) at composition $x=0.5$. (Left) Convensional CP in terms of second-order moment of $\bm{A}_1$, $\bm{A}_2$ and $\bm{A}_3$, corresponding to GIM correlation for 1-3NN pair. (Right) Extended CP based on the proposed graph spectrum, using graph energy for $\bm{A}_1$, $\bm{A}_2$ and $\bm{A}_3$. Structure at the vertices are emphasized by closed circles.  }
\label{fig:cp0.5}
\end{center}
\end{figure}

\begin{figure}[h]
\begin{center}
\includegraphics[width=0.95\linewidth]
{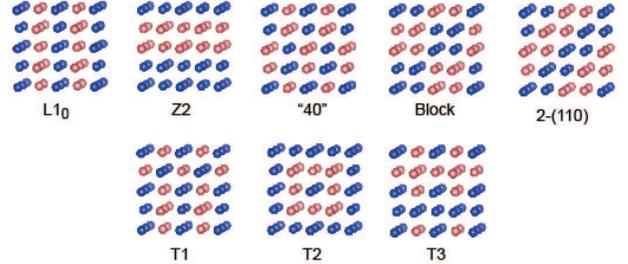}
\caption{ Atomic configuration of nine ordered structures that are found at vertices of convensional and/or extended CP in Fig.~\ref{fig:cp0.5}.  }
\label{fig:str0.5}
\end{center}
\end{figure}

Figure~\ref{fig:cp0.5} shows the constructed CP for convensional GIM (left) and extended graph (right) representation at $x=0.5$, in terms of adjacency matrix of 1-3 NN pairs. 
From Fig.~\ref{fig:cp0.5} (a), we can find five ordered structures of L1$_0$, "40", Z2, Block and 2-(110) at the vertices. L1$_0$ and "40" are known to be ground-state atomic configurations for real alloys of Cu-Au and Pt-Rh system,\cite{ord1, ord2} and Z2 of alternating two-layer stacking along (001) has been predicted as ground-state for Pt-Ru alloy based on systematic first-principles study\cite{ord3}): We can certainly confirm that structures at vertices of CP can be candidate for ground-state structure. 
When we construct the extended CP using linear combination of the same $\bm{A}$s, the resultant CP is found to have more vertices than the convensional one, as shown in 
Fig.~\ref{fig:cp0.5} (b). We can clearly see that three additional structures of T1, T2 and T3 are found at the extended CP, and five structures at vertices of convensional CP are all locates at those of the extended CP, which has been confirmed in our previous study. 
When we focus the lower two figures of Fig.~\ref{fig:cp0.5} for CP of 2 and 3NN pair figures, similar tendency can be found: Three ordered structures of L1$_0$, "40" and Z2 can be found at vertices of both convensional and extended CP, and two additional structures of T1 and T3 are also found at those of the extended CP. 
Figure~\ref{fig:str0.5} shows atomic configuration of the nine ordered structures at vertices of CPs in Fig.~\ref{fig:cp0.5}. 
These results certainly indicates that the proposed extended CP based on graph theory can not only capture a set of structure found at vertices of conventional CP, but also can find other charcteristic structures in terms of graph (i.e., having characteristic links). 

\begin{figure}[h]
\begin{center}
\includegraphics[width=1.00\linewidth]
{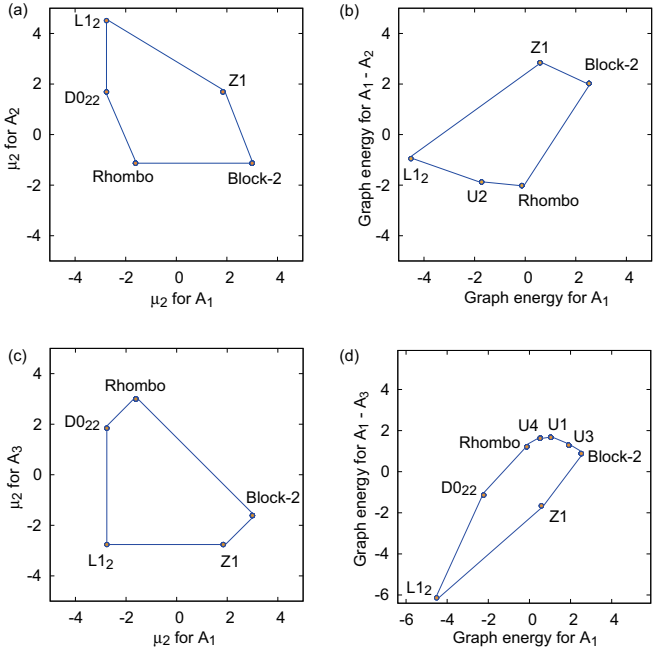}
\caption{Configurational polyhedra (CP) at composition $x=0.75$. (Left) Convensional CP in terms of second-order moment of $\bm{A}_1$, $\bm{A}_2$ and $\bm{A}_3$, corresponding to GIM correlation for 1-3NN pair. (Right) Extended CP based on the proposed graph spectrum, using graph energy for $\bm{A}_1$, $\bm{A}_2$ and $\bm{A}_3$. Structure at the vertices are emphasized by closed circles. }
\label{fig:cp0.75}
\end{center}
\end{figure}

\begin{figure}[h]
\begin{center}
\includegraphics[width=0.95\linewidth]
{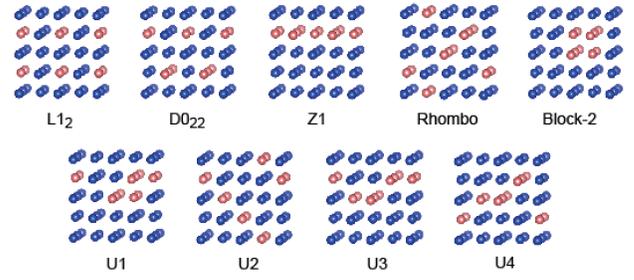}
\caption{Atomic configuration of nine ordered structures that are found at vertices of convensional and/or extended CP in Fig.~\ref{fig:cp0.75}.}
\label{fig:str0.75}
\end{center}
\end{figure}

In order to see the composition dependence of these tendencies, we also construct conventional and extended CPs at $x=0.75$, based on $\bm{A}_1$, $\bm{A}_2$ and $\bm{A}_3$. 
Figure~\ref{fig:cp0.75} shows the resultant CPs, in terms of adjacency matrix of 1-3 NN pairs. From Fig.~\ref{fig:cp0.75} (a) and (c), we find L1$_2$ and D0$_{22}$ ordered structures that have been 
considered ground states for real alloys, and additional three ordered structures of Rhombo, Z1 and Block-2 are found. 
Using the same $\bm{A}$s, the extended CPs in Fig.~\ref{fig:cp0.75} (b) and (d) successfully retains these five ordered structures at their vertices, and also have additional four ordered structures of U1, U2, U3 and U4. These atomic arrangements are illustrated in Fig.~\ref{fig:str0.75}.
Therefore, we can see that the proposed extended graph representation generally retains vertices of conventional CP as well as provides other characteristic ordered structures using the same set of pair figures on the same dimension of configuration space, which is a desired property as described above. 

The reason why the proposed representation can find out more characteristic structures than conventional one certainly reflects that the landscape of the proposed graph spectrum has more structural information (especially, higher-order structural links) that cannot be explicitly included in the GIM description. For instance, graph energy naturally include the information about asymmetric landscape of graph spectrum. When we quantitatively determine this asymmetry as third-order moment of the spectrum, $\mu_3$, this can be explicitly given by, for instance, 
\begin{widetext}
\begin{eqnarray}
\label{eq:3m}
&&\mu_3 \left[ \mathrm{Spec}\left(\bm{A}_1 - \bm{A}_3\right)  \right] = N^{-1}\sum_i\sum_j\sum_k \Braket{i|A_{1\overline{3}}|j} \Braket{j|A_{1\overline{3}}|k} \Braket{k|A_{1\overline{3}}|i} \nonumber \\
&&= \mu_3 \left[ \mathrm{Spec}\left(\bm{A}_1 \right)  \right]  - \mu_3 \left[ \mathrm{Spec}\left( \bm{A}_3\right)  \right]  + N^{-1}\left\{\sum_{i,j,k\in \left(133\right)} \Braket{i|A_{13}|j} \Braket{j|A_{13}|k} \Braket{k|A_{13}|i}  - \sum_{i,j,k\in \left(113\right)} \Braket{i|A_{13}|j} \Braket{j|A_{13}|k} \Braket{k|A_{13}|i} \right\}, \nonumber \\
\end{eqnarray}
\end{widetext}
where $\bm{A}_{13} = \bm{A}_1 + \bm{A}_3$ and $\bm{A}_{1\overline{3}} = \bm{A}_1 - \bm{A}_3$. Summation in the third term of right-hand equation is taken over symmetry-equivalent triplet composed of one  1NN and two 3NN pairs ($i,j,k\in\left(133\right)$) and of two 1NN and one 3NN pairs ($i,j,k\in\left(113\right)$). 

From the above equation, it is now clear that takeing a linear combination of $\bm{A}$s results in characterizing other higher-order links since $\mu_3\left[\mathrm{Spec} \left(\bm{A}_1 - \bm{A}_3\right)\right] \neq \mu_3 \left[ \mathrm{Spec}\left(\bm{A}_1 \right)  \right]  - \mu_3 \left[ \mathrm{Spec}\left( \bm{A}_3\right)  \right] $.
More explicitly, differences in asymmetry corresponds to difference in number of closed triplet links composed of A element (i.e., element of $\sigma = +1$) with one (two) 1NN and two (one) 3NN pairs. 
Graph energy not only contains information about third order moment, but also contains second-order moment (corresponding to GIM pair correlations) as well as higher-order moment (corresponding to higher-order structural links composed of $\sigma=+1$ element), which successfully results in the desired property found for the extended CP based on graph representation compared with the conventional one, as seen in Figs.~\ref{fig:cp0.5} and~\ref{fig:cp0.75}. 


\section{Conclusions}
Based on the graph theory and generalized Ising model, we propose theoretical approach to construct configurational polyhedra (CP) for crystalline solids. The extended CP have vertices not only include those found in conventional CP, but also include other characteristic structures on the same dimensional configuration space with the same set of figures composed of underlying lattice points, which therefore has significant advantage over the conventional approach. We confirm that this desired property can be naturally due to the fact that the proposed representation has more structural information, especially about higher-order structural links for selected element, than conventional GIM description.

\section*{Acknowledgement}
This work was supported by a Grant-in-Aid for Scientific Research (16K06704) from the MEXT of Japan, Research Grant from Hitachi Metals$\cdot$Materials Science Foundation, and Advanced Low Carbon Technology Research and Development Program of the Japan Science and Technology Agency (JST).

\end{document}